\documentclass[aip,jmp,amsmath,amssymb,preprint]{revtex4-1}
 \usepackage{graphicx}
 \usepackage{bm}
 \usepackage{bbm}

 \usepackage{subfigure}
 \usepackage{amssymb}

\usepackage{dcolumn}
\usepackage{bm}
\usepackage{xcolor}[2007/01/21]
\begin{document}

\title{Effect of nanostructure layout on spin pumping phenomena in antiferromagnet/ nonmagnetic metal/ ferromagnet multilayered stacks}

\author{A. F. Kravets}
\email{anatolii@kth.se}
\affiliation{Institute of Magnetism, National Academy of Sciences of Ukraine, 03680 Kyiv, Ukraine}
\affiliation{Nanostructure Physics, Royal Institute of Technology, 10691 Stockholm, Sweden}

\author{Olena V. Gomonay}
\affiliation{Institut f\"ur Physik, Johannes Gutenberg Universit\"at Mainz, D-55099 Mainz, Germany}
\affiliation{National Technical University of Ukraine ``KPI'', 03056, Kyiv, Ukraine}

\author{D. M. Polishchuk}
\author{Yu. O. Tykhonenko-Polishchuk}
\affiliation{Nanostructure Physics, Royal Institute of Technology, 10691 Stockholm, Sweden}
\affiliation{Institute of Magnetism, National Academy of Sciences of Ukraine, 03680 Kyiv, Ukraine}

\author{T. I. Polek}
\author{A. I. Tovstolytkin}
\affiliation{Institute of Magnetism, National Academy of Sciences of Ukraine, 03680 Kyiv, Ukraine}

\author{V. Korenivski}
\affiliation{Nanostructure Physics, Royal Institute of Technology, 10691 Stockholm, Sweden}

\date{\today}

\begin{abstract}

In this work we focus on magnetic relaxation in Mn$_{80}$Ir$_{20}$(12 nm)/ Cu(6 nm)/ Py($d_\mathrm{F}$) antiferromagnet/Cu/ferromagnet (AFM/Cu/FM) multilayers with different thickness of the ferromagnetic permalloy layer. An effective FM-AFM interaction mediated via the conduction electrons in the nonmagnetic Cu spacer -- the spin-pumping effect -- is detected as an increase in the linewidth of the ferromagnetic resonance (FMR) spectra and a shift of the resonant magnetic field. We further find experimentally that the spin-pumping-induced contribution to the linewidth is inversely proportional to the thickness of the Py layer. We show that this thickness dependence likely originates from the dissipative dynamics of the free and localized spins in the AFM layer. The results obtained could be used for tailoring the dissipative properties of spintronic devices incorporating antiferromagnetic layers.

\end{abstract}


\maketitle

Antiferromagnets (AFMs) are attractive materials for spintronic applications. They operate at high frequencies and thus have the potential to functionally fill the ``terahertz gap'' in electronics. Due to their lack of a macroscopic magnetic moment, AFMs produce no stray fields and therefore potentially can provide higher scalability for magnetic memory devices. High typical values of the spin-flop fields prevent AFMs from spontaneous thermally-induced switching and increase the data retention times. In addition, recent experimental\cite{MacDonald2011} and theoretical investigations\cite{Gomonay2014} have shown that AFMs are sensitive to spin-polarized currents and can be used as active elements in spintronic devices.

Direct observation of spintronic effects in AFMs is challenging due precisely to the same reasons that make AFMs competitive with their ferromagnetic counterparts: the magnetoresistance in AFM-based devices is low due to the absence of net magnetization in AFM, and the dynamics require very high excitation frequencies, beyond the capabilities of microwave circuits. An alternative technique to detect the spin dynamics of AFM films was recently implemented by a number of groups.\cite{Merodio2014,Frangou2016,Wang2014,Hahn2014,Qiu2016} This technique is based on the spin pumping effect, which is reciprocal to the spin-transfer torque effect.\cite{Tserkovnyak2002,Brataas2012} A metallic ferromagnetic layer (FM) is excited at its resonance frequency (FMR) and pumps spin current into a neighbouring non-magnetic layer interfaced with an antiferromagnetic film (AFM) at the other surface. The linewidth of the FMR spectrum increases due to the presence of the AFM layer and thereby provides information about the interaction of the nonequilibrium conduction-electron spins and the localized AFM moments. 

The interpretation of such experiments is not quite straightforward, however, as different processes contribute to the effective damping in a multilayered sample: spin-dependent scattering at the interfaces \cite{Rezende2003} and in the bulk, energy exchange between the free and localised spins, spin-diffusion, etc. An efficient theoretical approach to this problem, based on nonequilibrium thermodynamics, was proposed in Ref.~\onlinecite{Brataas2006} for ferromagnetic (FM)/nonmagnetic (NM) bilayers, and was further generalized for FM/NM/FM systems.\cite{Taniguchi2014} Spin-pumping from an AFM layer was recently predicted in Refs.~\onlinecite{Cheng2014,Zhang2014}.

In this paper we focus on the dissipative response, expressed via the FMR linewidth, of MnIr/Cu/Py multilayers with different thickness of the Py layer. We generalize the Onsager formalism for the case of the discrete system AFM/NM/FM and calculate the effective Gilbert damping of the FM layer, taking into account the spin-pumping and spin-accumulation effects in both the FM and AFM layers. While the previous experiments\cite{Merodio2014,Frangou2016} have studied the damping dependence vs thickness of the AFM layer, we focus on the properties of the FM layer and especially the FM/NM interface. Our experiments reveal an inverse dependence of the additional, AFM-induced damping on the thickness of the FM layer, in agreement with our theoretical predictions. Our results should be useful for tailoring dissipation in spintronic devices. 

For the experiments we use multilayers Substrate/Ta(5)/Py(3)/Mn$_{80}$Ir$_{20}$(12)/Cu(6)/ Py($d_\mathrm{F}$)/Al(4), hereinafter  AFM/Cu/FM($d_\mathrm{F}$), with the FM layer of variable thickness, $d_\mathrm{F}=$ 3, 6, 9, 12, 15 nm. The numbers in parenthesis denote thickness in nanometers of the corresponding layers; Py = Ni$_{80}$Fe$_{20}$. In these multilayers, Mn$_{80}$Ir$_{20}$(12), Cu(6) and Py($d_\mathrm{F}$) form the functional combination of the AFM/NM/FM stack, while the other layers are auxiliary. The top Al layer is a protective capping layer. The bottom layers facilitate the formation of the optimal crystalline and magnetic structure of Mn$_{80}$Ir$_{20}$(12). We also fabricated a set of reference samples with identical structure but without Py(3)/Mn$_{80}$Ir$_{20}$(12) layers.

The multilayers were deposited at room temperature (295 K) on thermally oxidized silicon substrates using magnetron sputtering in an AJA Orion 8-target system.\cite{Kravets2012} The base pressure in the deposition chamber was 5 $\times 10^{−8}$~Torr and the Ar pressure used during deposition was 3 mTorr. The exchange pinning between Py(3) and Mn$_{80}$Ir$_{20}$(12) layers was set in during the deposition of the multilayers using an in-plane magnetic field of 1 kOe.

\begin{figure}
\centering
\includegraphics[width=0.9\linewidth]{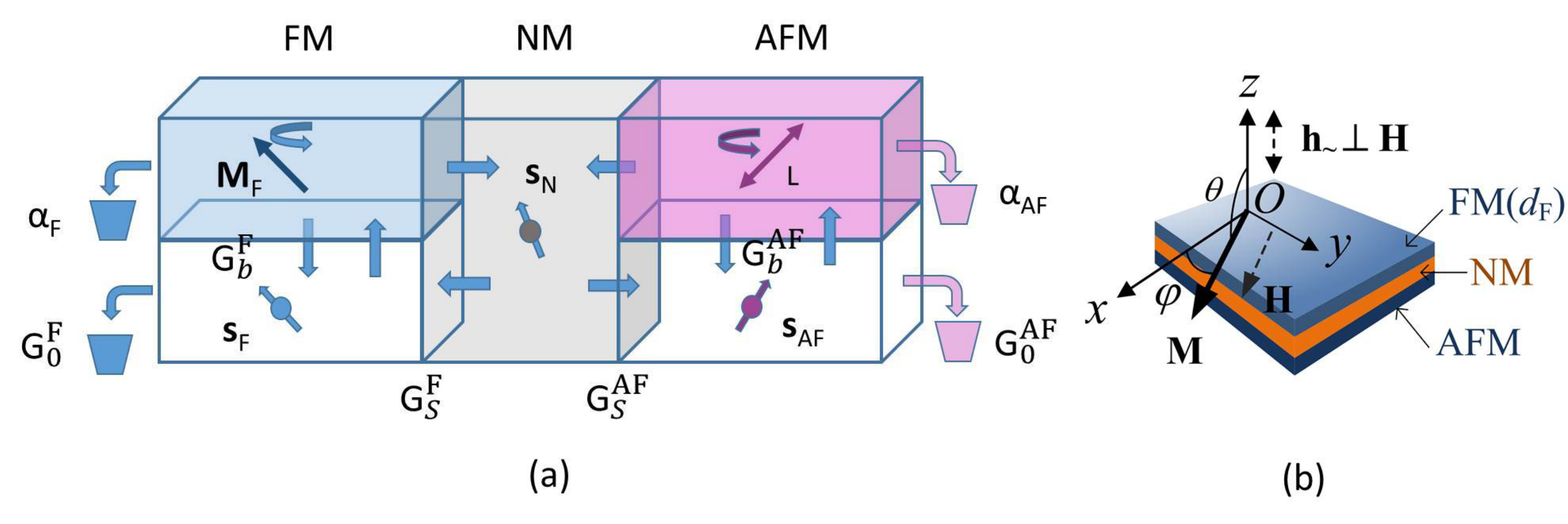}
\caption{(a) Schematic view of the energy and spin exchange within a trilayer system FM/NM/AFM. The magnetic layers (FM and AFM) are symbolically separated into subsystems of localized (coloured area) and free (white area) spins. Wide arrows show the fluxes that originate from different mechanisms. Vertical arrows correspond to spin exchange between the localized and free spins inside the FM and AFM layers. (b) Schematic view of the FMR experiment, where $\textbf{H}\parallel xOy$ is the in-plane geometry and $\textbf{H}\parallel xOz$ is the out-of-plane geometry.}
\label{scheme}
\end{figure}

\begin{figure}
\centering
\includegraphics[width=\linewidth]{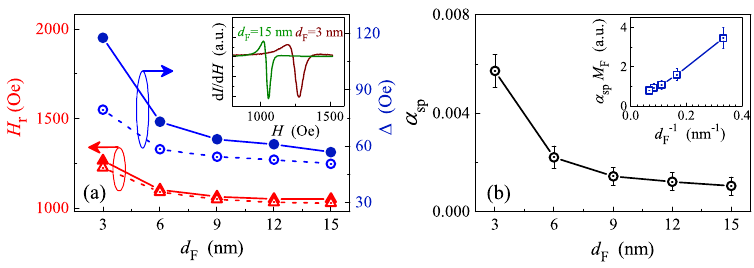}
\caption{(a)~In-plane resonance field $H_\mathrm{r}$ (triangles) and linewidth $\Delta$ (circles) vs thickness $d_\mathrm{F}$ of the Py layer for AFM/Cu/FM($d_\mathrm{F}$) multilayers (bold symbols, solid lines) and for  reference samples (open symbols, dashed lines). Inset shows typical FMR spectra for AFM/Cu/FM($d_\mathrm{F}$) samples with $d_\mathrm{F}=$ 3 and 15 nm. (b) Py-thickness dependence of $\alpha_\mathrm{sp}$ for AFM/Cu/FM($d_\mathrm{F}$). Inset shows $\alpha_\mathrm{sp} M_{\mathrm{F}}$ product as a function of $d_\mathrm{F}^{-1}$. The solid line is guide to the eye.}
\label{experiment}
\end{figure}

We use an X-band ELEXSYS E500 spectrometer equipped with an automatic goniometer to measure the out-of-plane and in-plane angular dependencies of the FMR spectra. The operating frequency is 9.85 GHz, the temperature is 295 K. The spectra show no signal from the Py(3) buffer layer, while the signal from Py($d_\mathrm{F}$) is clearly visible. We record the magnetic-field derivative of the microwave absorption and fit each spectrum by a Lorentzian function to obtain the resonance field $H_\mathrm{r}$ and the linewidth $\Delta$ in the in-plane and the out-of plane geometries [Fig.~\ref{scheme}(b)]. Typical FMR spectra measured for the in-plane orientation are shown in the inset to Fig.~\ref{experiment}(a).

When FMR is excited, a moving magnetization in the FM pumps a spin current into the NM and AFM layers.\cite{Tserkovnyak2005} The spin current is proportional to the effective field $\mathbf{H}_\mathrm{F}$, which determines the magnetic dynamics in the FM layer. The spin current can induce exchange of angular momentum between the different subsystems of the conduction and localized electrons in the NM and AFM layers. Moreover, it can stimulate additional spin pumping from the AFM layer induced by the dynamic magnetization $\mathbf{M}_\mathrm{AF}$, which follows the motion of the localized AFM moments.\cite{Baryakhtar1979,Gomonay2008,Cheng2014} In addition, free conduction-electron spins in our metallic AFM can interact with the dynamic magnetization $\mathbf{M}_\mathrm{AF}$ and also accumulate, similar to that in the NM layer. While the spin polarization in FM is so strong that spin accumulation in it can be neglected, in the metallic AFM spin accumulation and spin polarization by the localized moments are comparable. Therefore, the transport of spins through the AFM/NM/FM system and the corresponding dissipative phenomena within the trilayer depend upon the balance between the free and localized spins within all three layers of the structure. 

Treating the AFM/NM/FM as a discrete system, one can distinguish between five subsystems, shown schematically in Fig.~\ref{scheme}(a): three reservoirs of free spins in FM (spin density $\mathbf{s}_\mathrm{F}$), NM (spin density $\mathbf{s}_\mathrm{N}$), and AFM (spin density $\mathbf{s}_\mathrm{AF}$), and localized FM (macroscopic magnetization $\mathbf{M}_\mathrm{F}\equiv M_\mathrm{F}\mathbf{m}_\mathrm{F}$) and AFM moments (characterized with the N\'eel order parameter $\mathbf{L}=M_\mathrm{AF}\mathbf{l}$ and macroscopic magnetization $\mathbf{M}_\mathrm{AF}\equiv M_\mathrm{AF}\mathbf{m}_\mathrm{AF}$). Here we introduce the saturation magnetizations $M_\mathrm{F}$  and $M_\mathrm{AF}$ of the FM and AFM layers, respectively. In equilibrium, free spins in the FM are mostly parallel to the FM magnetization, $\mathbf{s}_\mathrm{F}\|\mathbf{M}_\mathrm{F}$. In the NM and AFM layers, the population of free spin-up and spin-down electrons is balanced, $\mathbf{s}_\mathrm{N}=\mathbf{s}_\mathrm{AF}=0$, since $\mathbf{M}_\mathrm{AF}=0$.

In the framework of linear nonequilibrium thermodynamics, spin densities $\mathbf{s}_\mathrm{F}$, $\mathbf{s}_\mathrm{N}$, $\mathbf{s}_\mathrm{AF}$, and magnetizations $\mathbf{m}_\mathrm{F}$, $\mathbf{m}_\mathrm{AF}$ can be treated as thermodynamic variables $a_j$, $j=1\ldots 5$. The conjugated thermodynamic forces are calculated as the derivatives of free energy:\cite{Callen1948} $X_j=\partial F/\partial a_j$ (we assume that the temperature is constant). The thermodynamic forces for the free spins coincide with the spin accumulation potentials $\boldsymbol{\mu}^{(s)}_\mathrm{F}$ (in FM), $\boldsymbol{\mu}^{(s)}_\mathrm{N}$ (in NM), and $\boldsymbol{\mu}^{(s)}_\mathrm{AF}$ (in AFM). For the localized moments the corresponding forces are proportional to the effective fields $M_\mathrm{F}V_\mathrm{F}\mathbf{H}_\mathrm{F}$ (in FM) and  $M_\mathrm{AF}V_\mathrm{AF}\mathbf{H}_\mathrm{AF}$ (in AFM).

Thermodynamic currents $J_j\equiv \dot{a}_j$ are related to the thermodynamic forces via the Onsager coefficients $\hat {\mathcal{L}}$: 
\begin{equation}\label{eq_general_relation}
\left(\dot{\mathbf{m}}_\mathrm{AF} ,\dot{\mathbf{m}}_\mathrm{F} ,\dot{\mathbf{s}}_\mathrm{AF}/e,\dot{\mathbf{s}}_\mathrm{F}/e,\dot{\mathbf{s}}_\mathrm{N}/e \right)^\mathrm{T}=\hat {\mathcal{L}}\left(M_\mathrm{AF}V_\mathrm{AF}\mathbf{H}_\mathrm{AF} ,M_\mathrm{F}V_\mathrm{F}\mathbf{H}_\mathrm{F},\boldsymbol{\mu}^{(s)}_\mathrm{AF},\boldsymbol{\mu}^{(s)}_\mathrm{F} ,\boldsymbol{\mu}^{(s)}_\mathrm{N}   \right)^\mathrm{T},
\end{equation}
where $e$ is electron charge. 

Using the Onsager reciprocity principle and the symmetry considerations, one can reduce relations (\ref{eq_general_relation}) to the following form:
\begin{eqnarray}  \label{eq_Onsager_relations_1}
\dot{\mathbf{m}}_\mathrm{AF}&=&\gamma\alpha_\mathrm{AF}\mathbf{H}_\mathrm{AF}%
-\frac{\gamma\hbar }{e^2M_\mathrm{AF}V_\mathrm{AF}}\mathbf{l}\times
\left(G^\mathrm{AF}_\mathrm{b}\boldsymbol{\mu}^{(s)}_%
\mathrm{AF}+G^\mathrm{AF}_\mathrm{S}{\mu}^{(s)}_\mathrm{F}\mathbf{m}_\mathrm{%
	F}\right)\times\mathbf{l},  \nonumber \\
\dot{\mathbf{m}}_\mathrm{F}&=&\gamma\alpha_\mathrm{F}\mathbf{H}_\mathrm{F}-%
\frac{\gamma\hbar G^\mathrm{F}_\mathrm{S} }{e^2M_\mathrm{F}V_\mathrm{F}}%
\mathbf{m}_\mathrm{F}\times
\boldsymbol{\mu}^{(s)}_\mathrm{AF}\times\mathbf{m}_\mathrm{F}, \\
\dot{\mathbf{s}}_\mathrm{AF}&=&-\frac{\gamma\hbar }{e}G^\mathrm{AF}_\mathrm{b%
}\mathbf{H}_\mathrm{AF}+\frac{1}{e}G^\mathrm{AF}_\mathrm{0}\boldsymbol{\mu}^{(s)}_%
\mathrm{AF}, \quad \dot{\mathbf{s}}_\mathrm{F}=-\frac{\gamma\hbar }{e}G^\mathrm{F}_\mathrm{b%
}\mathbf{H}_\mathrm{F}+\frac{1}{e}G^\mathrm{F}_\mathrm{0}{\mu}^{(s)}_\mathrm{F}\mathbf{m}_\mathrm{%
F},   \nonumber \\
\dot{\mathbf{s}}_\mathrm{N}&=&\frac{\gamma\hbar }{e}G^\mathrm{AF}_\mathrm{S}%
\mathbf{H}_\mathrm{AF}-\frac{\gamma\hbar }{e}G^\mathrm{F}_\mathrm{S}\mathbf{%
	H}_\mathrm{F}+\frac{1}{e}G^\mathrm{N}(\boldsymbol{\mu}%
^{(s)}_\mathrm{F} -\boldsymbol{\mu}%
^{(s)}_\mathrm{AF}) ,  \nonumber
\nonumber
\end{eqnarray}
where $\gamma$ is the gyromagnetic ratio and $\hbar$ is the Plank constant. We neglect spin accumulations in the NM layer, since the spin-diffusion length in the NM layer is relatively long. We also set $\boldsymbol{\mu}^{(s)}_\mathrm{N}=0$ and take into account strong spin polarization in the FM layer, so, that $\boldsymbol{\mu}^{(s)}_ \mathrm{F}={\mu}^{(s)}_\mathrm{F}\mathbf{m}_\mathrm{F}$. In the second equation of (\ref{eq_Onsager_relations_1}) we use Landau-Lifshitz representation of the magnetic damping in FM ($\propto \alpha_\mathrm{F}\mathbf{H}_\mathrm{F}$), as it is consistent with the Onsager's concept of conjugated currents ($\dot{\mathbf{m}}_\mathrm{F}$) and forces ($\mathbf{H}_\mathrm{F}$). Conversion to the standard Gilbert form can be obtained from  equations of motions for FM as $\mathbf{H}_\mathrm{F}=\mathbf{m}_\mathrm{F}\times\dot{\mathbf{m}}_\mathrm{F}/\gamma$.

The interpretation of the coefficients in Eq.~(\ref{eq_Onsager_relations_1}) is schematically shown in Fig.~\ref{scheme}(a). Diagonal coefficients $\mathcal{L}_{jj}$ for the localized spins are related with the internal damping in the FM (damping parameter $\alpha_\mathrm{F}$) and AFM (damping parameter $\alpha_\mathrm{AF}$) layers. Diagonal coefficients $\mathcal{L}_{jj}$ for the free spins are proportional to the corresponding conductances, $G^\mathrm{F}_\mathrm{0}$ and $G^\mathrm{AF}_\mathrm{0}$. The nondiagonal coefficients responsible for the cross-coupling effects between the AFM and FM layers, are of two types. First, the spin-mixing conductances $G^\mathrm{F}_\mathrm{S}$ and $G^\mathrm{AF}_\mathrm{S} $ originate from the dephasing of the free electrons at the FM/NM and NM/AFM interfaces,\cite{Tserkovnyak2002} and are responsible for the spin-pumping phenomena. The free electrons in NM reflecting from the FM/NM and NM/AFM interfaces acquire additional nonequilibrium spin polarization, which is related to the dynamic magnetization of the FM and AFM films. Second, the bulk conductivities, $G^\mathrm{AF}_\mathrm{b}$ and $G^\mathrm{F}_\mathrm{b}$ describe the exchange of angular momentum between the subsystems of the localized and free spins in the FM and AFM layers. In our case of strong polarization inside the FM layer, the term with $G^\mathrm{F}_\mathrm{b}$ can be neglected. Lastly, $G^\mathrm{N}$ is the spin conductivity in the NM layer. 

The first of Eqs.~(\ref{eq_Onsager_relations_1}) reproduces the well-known result of AFM spintronics:\cite{Gomonay2014,Cheng2014a} the spin-torque induced by a spin-polarized current (last term in the r.h.s.). It is clear from Eq.~(\ref{eq_Onsager_relations_1}) that this torque originates not only from the current polarized by the FM layer, but also from the spin accumulation inside the AFM layer.    

The second of Eqs.~(\ref{eq_Onsager_relations_1}), for the FM magnetization is similar to the corresponding equation for FM/NM bilayers,\cite{Brataas2011,Brataas2012,Ohnuma2014,Frangou2016} with the only difference that spin accumulation $\boldsymbol{\mu}^{(s)}_\mathrm{AF}$ takes place in the AFM layer. This fact reflects the ``duality'' of the metallic AFM, which manifests the properties of FM (non-zero magnetization of the localized spins) as well as NM (has free spins that can accumulate). There is one principal difference, however, between spin accumulation $\boldsymbol{\mu}^{(s)}_\mathrm{F}$ in FM/NM systems, and $\boldsymbol{\mu}^{(s)}_\mathrm{AF}$ in AFM/NM/FM trilayers. The FM magnetization is large and fully defines the orientation of the spin accumulation in a FM/NM bilayer. In the AFM/NM/FM system, the spin accumulation $\boldsymbol{\mu}^{(s)}_\mathrm{AF}$  is defined by the interplay between the magnetic dynamics in the AFM and the spin flow between the FM, NM, and AFM layers, with the result that its spin orientation is defined by a non-trivial interplay of a number of factors and point essentially in any direction.  

To describe the magnetic dynamics of a AFM/NM/FM trilayer one must start from the balance equations for the localized moments in the FM and AFM layers, and take into account the spin flows through the interfaces and the dissipative terms given by Eq.~(\ref{eq_Onsager_relations_1}). In particular, the equation for the FM moments can be written as
\begin{eqnarray}\label{eq_magnetic_FM}
\dot{\mathbf{m}}_\mathrm{F}&=&-\gamma \mathbf{m}_\mathrm{F}\times (\mathbf{H}_
\mathrm{F}+\mathbf{H})-\frac{1}{eM_\mathrm{F}V_\mathrm{F}}%
\mathbf{m}_\mathrm{F}\times\dot{\mathbf{s}}_\mathrm{N}\times\mathbf{m}_\mathrm{F}\nonumber\\
&+&\gamma\alpha_\mathrm{F}\mathbf{H}_\mathrm{F}-%
\frac{\gamma\hbar G^\mathrm{F}_\mathrm{S} }{e^2M_\mathrm{F}V_\mathrm{F}}%
\mathbf{m}_\mathrm{F}\times\boldsymbol{\mu}^{(s)}_\mathrm{AF}\times\mathbf{m}_\mathrm{F},
\end{eqnarray}

The first term in Eq.~(\ref{eq_magnetic_FM}) corresponds to the standard Landau-Lifshits dynamics in the presence of external magnetic field $\mathbf{H}$. The second term describes a spin flux through the interface, which coincides with the spin current, $-\dot{\mathbf{s}}_\mathrm{N}$, from the adjacent NM layer. Cross products with $\mathbf{m}_\mathrm{F}$ reflect the fact that only transverse (with respect to $\mathbf{m}_\mathrm{F}$) spin component flows out of FM. Last two terms correspond to the Onsager forces, according to Eq.~(\ref{eq_Onsager_relations_1}).

For the AFM/Cu/FM($d_\mathrm{F}$) system used in our FMR-induced spin pumping experiment, we can set $\boldsymbol{\mu}^{(s)}_\mathrm{F}=0$ as no electric voltage is applied across the structure. We further assume no spin accumulation inside the AFM layer, $\boldsymbol{\mu}^{(s)}_\mathrm{AF}=0$, since the spin-diffusion length in AFM (0.3 nm for Cu/IrMn \cite{Acharyya2011}) is much shorter than the AFM thickness. Then, from Eq.~(\ref{eq_general_relation}) and (\ref{eq_magnetic_FM}) we obtain the effective dynamic equation for the FM layer:
\begin{eqnarray}  \label{eq_reduced_FM_AFM}
\dot{\mathbf{m}}_\mathrm{F}&=&-\gamma \mathbf{m}_\mathrm{F}\times (\mathbf{H}_
\mathrm{F}+\mathbf{H})+\gamma\left(\alpha_\mathrm{F}+\frac{\gamma
	\hbar G^\mathrm{F}_\mathrm{S}}{e^2M_\mathrm{F}V_\mathrm{F}}\right)\mathbf{H}_\mathrm{F}+\frac{\gamma^2 \hbar^2 G^\mathrm{AF}_\mathrm{S}}{e^2M_%
	\mathrm{F}V_\mathrm{F}}%
\mathbf{m}_\mathrm{F}\times\mathbf{H}_\mathrm{AF}\times\mathbf{m}_%
\mathrm{F}.
\end{eqnarray}

The second term in the r.h.s. of Eq.~(\ref{eq_reduced_FM_AFM}) points to an increase of the effective damping due to the presence of the FM/NM interface, which leads to a corresponding increase in the FMR linewidth $\Delta$. In addition, the last term in Eq.~(\ref{eq_reduced_FM_AFM}) predicts a field-like contribution to the FM dynamics, which results exclusively from the spin pumping by the AFM layer, as the direct exchange between the FM and AFM is fully suppressed by the Cu spacer. This field, $\propto \mathbf{H}_\mathrm{AF}\times\mathbf{m}_\mathrm{F}$, can contribute to the value of the resonant field $H_r$, and the contribution can be estimated as follows. The typical AFMR frequencies are much larger than the FMR frequency of the FM layer, so the dynamics of the AFM is driven solely by the FM, and $\mathbf{H}_\mathrm{AF}\propto \mathbf{H}_\mathrm{F}$. The additional field is then $\propto G^\mathrm{AF}_\mathrm{S}\mathbf{H}_\mathrm{F}\times\mathbf{m}_\mathrm{F}/M_\mathrm{F}V_\mathrm{F}$.

According to Eq.~(\ref{eq_reduced_FM_AFM}), both spin-pumping-induced corrections to the linewidth and the resonant field are inversely proportional to $M_\mathrm{F}V_\mathrm{F}\propto M_\mathrm{F}d_\mathrm{F}$. Fig.~\ref{experiment}(a) illustrates this tendency of $\Delta(d_\mathrm{F})$ and $H_\mathrm{r}(d_\mathrm{F})$ measured for our samples.

To confirm the thickness dependence of the effective damping predicted by Eq.~(\ref{eq_reduced_FM_AFM}), we calculated the incremental change in the AFM-induced linewidth as $\Delta_\mathrm{sp} = \Delta - \Delta_\mathrm{inhom} - \Delta_\mathrm{ref}$, where $\Delta_\mathrm{ref}$ is the linewidth of the FMR of the reference sample. Contribution $\Delta_\mathrm{inhom}$, which originates from a possible inhomogeneuity of the sample is calculated according to the procedure described in Refs.~\onlinecite{Mizukami2001a,Kravets2016}. This contribution is below 2 Oe for the samples with $d_\mathrm{F}\leq 6$ nm and equals to 8 Oe for $d_\mathrm{F}=3$ nm. It should be noted that for the multilayer with the thickest Py layer ($d_\mathrm{F}=15$ nm), the FMR linewidth ($\sim 55$ Oe) well agrees with the values reported by other research groups for well-characterized high-quality Py films.\cite{Merodio2014,Montoya2016}

Figure~\ref{experiment}(b) shows the thickness dependence of the spin-pumping-induced contribution to Gilbert damping obtained from $\Delta_\mathrm{sp}$. In agreement with the theory, Eq.~(\ref{eq_reduced_FM_AFM}), $\alpha_\mathrm{sp}M_\mathrm{F}$ grows linearly with $d_\mathrm{F}^{-1}$. We believe that the observed thickness dependence of the damping parameter points to the important role of free spins in the magnetic dynamics of the AFM/Cu/FM trilayer. We also conclude that the observed $H_\mathrm{r}(d_\mathrm{F})$ dependence indicates that the localized AFM moments affect the dynamics of the FM layer through the dynamic exchange via conduction electrons in the system. However, this contribution from the localized moments can be partially  masked by the exchanges bias due to the second Py layer and thus requires further analysis. 

In summary, we observe spin-pumping effect in AFM/NM/FM multilayers as an increase in the linewidth of the FMR and shift of the resonant magnetic field. Basing on Onsager formalism, we calculate additional damping and field-like torque on FM moments due to the presence of AFM layer. The inverse dependence of damping and resonant field vs the thickness of FM layer supports the hypothesis of AFM influence on FM dynamics. The contribution from the spin-pumping effect to the FMR linewidth is separated and shown be affected by the changes in the thickness of the ferromagnetic layer. The physical mechanisms of the observed $\Delta_{\mathrm{sp}}$ vs. $d_\mathrm{F}$ behaviour are analyzed and show a rich interplay of the conduction-vs-lattice spins in the five effective sub-systems of the structure. These results provide a deeper understanding of the spintronic effects in nanostructures containing antiferromagnets and can prove useful for designing future spintronic devices.

\begin{acknowledgments}
Support from the Swedish Stiftelse Olle Engkvist Byggm\"{a}stare, the Swedish Research Council (VR grant 2014-4548), and the National Academy of Sciences of Ukraine (project 0115U00974) are gratefully acknowledged. OG acknowledges support of the Fundamental research program of the National Academy of Sciences of Ukraine ``Fundamental problems of creation of new nanomaterials and nanotechnologies'', the ERC Synergy Grant SC2 (No. 610115), and the Transregional Collaborative Research Center (SFB/TRR) 173 SPIN+X.
\end{acknowledgments}


\begin{thebibliography}{26}%
	\makeatletter
	\providecommand \@ifxundefined [1]{%
		\@ifx{#1\undefined}
	}%
	\providecommand \@ifnum [1]{%
		\ifnum #1\expandafter \@firstoftwo
		\else \expandafter \@secondoftwo
		\fi
	}%
	\providecommand \@ifx [1]{%
		\ifx #1\expandafter \@firstoftwo
		\else \expandafter \@secondoftwo
		\fi
	}%
	\providecommand \natexlab [1]{#1}%
	\providecommand \enquote  [1]{``#1''}%
	\providecommand \bibnamefont  [1]{#1}%
	\providecommand \bibfnamefont [1]{#1}%
	\providecommand \citenamefont [1]{#1}%
	\providecommand \href@noop [0]{\@secondoftwo}%
	\providecommand \href [0]{\begingroup \@sanitize@url \@href}%
	\providecommand \@href[1]{\@@startlink{#1}\@@href}%
	\providecommand \@@href[1]{\endgroup#1\@@endlink}%
	\providecommand \@sanitize@url [0]{\catcode `\\12\catcode `\$12\catcode
		`\&12\catcode `\#12\catcode `\^12\catcode `\_12\catcode `\%12\relax}%
	\providecommand \@@startlink[1]{}%
	\providecommand \@@endlink[0]{}%
	\providecommand \url  [0]{\begingroup\@sanitize@url \@url }%
	\providecommand \@url [1]{\endgroup\@href {#1}{\urlprefix }}%
	\providecommand \urlprefix  [0]{URL }%
	\providecommand \Eprint [0]{\href }%
	\providecommand \doibase [0]{http://dx.doi.org/}%
	\providecommand \selectlanguage [0]{\@gobble}%
	\providecommand \bibinfo  [0]{\@secondoftwo}%
	\providecommand \bibfield  [0]{\@secondoftwo}%
	\providecommand \translation [1]{[#1]}%
	\providecommand \BibitemOpen [0]{}%
	\providecommand \bibitemStop [0]{}%
	\providecommand \bibitemNoStop [0]{.\EOS\space}%
	\providecommand \EOS [0]{\spacefactor3000\relax}%
	\providecommand \BibitemShut  [1]{\csname bibitem#1\endcsname}%
	\let\auto@bib@innerbib\@empty
	\bibitem [{\citenamefont {MacDonald}\ and\ \citenamefont
		{Tsoi}(2011)}]{MacDonald2011}%
	\BibitemOpen
	\bibfield  {author} {\bibinfo {author} {\bibfnamefont {A.~H.}\ \bibnamefont
			{MacDonald}}\ and\ \bibinfo {author} {\bibfnamefont {M.}~\bibnamefont
			{Tsoi}},\ }\href {\doibase 10.1098/rsta.2011.0014} {\bibfield  {journal}
		{\bibinfo  {journal} {Phil. Trans. R. Soc. A}\ }\textbf {\bibinfo {volume}
			{369}},\ \bibinfo {pages} {3098} (\bibinfo {year} {2011})}\BibitemShut
	{NoStop}%
	\bibitem [{\citenamefont {Gomonay}\ and\ \citenamefont
		{Loktev}(2014)}]{Gomonay2014}%
	\BibitemOpen
	\bibfield  {author} {\bibinfo {author} {\bibfnamefont {E.~V.}\ \bibnamefont
			{Gomonay}}\ and\ \bibinfo {author} {\bibfnamefont {V.~M.}\ \bibnamefont
			{Loktev}},\ }\href {\doibase http://dx.doi.org/10.1063/1.4862467} {\bibfield
		{journal} {\bibinfo  {journal} {Low Temp. Phys.}\ }\textbf {\bibinfo {volume}
			{40}},\ \bibinfo {pages} {17} (\bibinfo {year} {2014})}\BibitemShut {NoStop}%
	\bibitem [{\citenamefont {Merodio}\ \emph {et~al.}(2014)\citenamefont
		{Merodio}, \citenamefont {Ghosh}, \citenamefont {Lemonias}, \citenamefont
		{Gautier}, \citenamefont {Ebels}, \citenamefont {Chshiev}, \citenamefont
		{B\'{e}a}, \citenamefont {Baltz},\ and\ \citenamefont
		{Bailey}}]{Merodio2014}%
	\BibitemOpen
	\bibfield  {author} {\bibinfo {author} {\bibfnamefont {P.}~\bibnamefont
			{Merodio}}, \bibinfo {author} {\bibfnamefont {A.}~\bibnamefont {Ghosh}},
		\bibinfo {author} {\bibfnamefont {C.}~\bibnamefont {Lemonias}}, \bibinfo
		{author} {\bibfnamefont {E.}~\bibnamefont {Gautier}}, \bibinfo {author}
		{\bibfnamefont {U.}~\bibnamefont {Ebels}}, \bibinfo {author} {\bibfnamefont
			{M.}~\bibnamefont {Chshiev}}, \bibinfo {author} {\bibfnamefont
			{H.}~\bibnamefont {B\'{e}a}}, \bibinfo {author} {\bibfnamefont
			{V.}~\bibnamefont {Baltz}}, \ and\ \bibinfo {author} {\bibfnamefont {W.~E.}\
			\bibnamefont {Bailey}},\ }\href {\doibase
		http://dx.doi.org/10.1063/1.4862971} {\bibfield  {journal} {\bibinfo
			{journal} {Appl. Phys. Lett.}\ }\textbf {\bibinfo {volume} {104}},\ \bibinfo
		{eid} {032406} (\bibinfo {year} {2014})}\BibitemShut {NoStop}%
	\bibitem [{\citenamefont {Frangou}\ \emph {et~al.}(2016)\citenamefont
		{Frangou}, \citenamefont {Oyarz\'un}, \citenamefont {Auffret}, \citenamefont
		{Vila}, \citenamefont {Gambarelli},\ and\ \citenamefont
		{Baltz}}]{Frangou2016}%
	\BibitemOpen
	\bibfield  {author} {\bibinfo {author} {\bibfnamefont {L.}~\bibnamefont
			{Frangou}}, \bibinfo {author} {\bibfnamefont {S.}~\bibnamefont {Oyarz\'un}},
		\bibinfo {author} {\bibfnamefont {S.}~\bibnamefont {Auffret}}, \bibinfo
		{author} {\bibfnamefont {L.}~\bibnamefont {Vila}}, \bibinfo {author}
		{\bibfnamefont {S.}~\bibnamefont {Gambarelli}}, \ and\ \bibinfo {author}
		{\bibfnamefont {V.}~\bibnamefont {Baltz}},\ }\href {\doibase
		10.1103/PhysRevLett.116.077203} {\bibfield  {journal} {\bibinfo  {journal}
			{Phys. Rev. Lett.}\ }\textbf {\bibinfo {volume} {116}},\ \bibinfo {pages}
		{077203} (\bibinfo {year} {2016})}\BibitemShut {NoStop}%
	\bibitem [{\citenamefont {Wang}\ \emph {et~al.}(2014)\citenamefont {Wang},
		\citenamefont {Du}, \citenamefont {Hammel},\ and\ \citenamefont
		{Yang}}]{Wang2014}%
	\BibitemOpen
	\bibfield  {author} {\bibinfo {author} {\bibfnamefont {H.}~\bibnamefont
			{Wang}}, \bibinfo {author} {\bibfnamefont {C.}~\bibnamefont {Du}}, \bibinfo
		{author} {\bibfnamefont {P.~C.}\ \bibnamefont {Hammel}}, \ and\ \bibinfo
		{author} {\bibfnamefont {F.}~\bibnamefont {Yang}},\ }\href {\doibase
		10.1103/PhysRevLett.113.097202} {\bibfield  {journal} {\bibinfo  {journal}
			{Phys. Rev. Lett.}\ }\textbf {\bibinfo {volume} {113}},\ \bibinfo {pages}
		{097202} (\bibinfo {year} {2014})}\BibitemShut {NoStop}%
	\bibitem [{\citenamefont {Hahn}\ \emph {et~al.}(2014)\citenamefont {Hahn},
		\citenamefont {de~Loubens}, \citenamefont {Naletov}, \citenamefont
		{Ben~Youssef}, \citenamefont {Klein},\ and\ \citenamefont
		{Viret}}]{Hahn2014}%
	\BibitemOpen
	\bibfield  {author} {\bibinfo {author} {\bibfnamefont {C.}~\bibnamefont
			{Hahn}}, \bibinfo {author} {\bibfnamefont {G.}~\bibnamefont {de~Loubens}},
		\bibinfo {author} {\bibfnamefont {V.~V.}\ \bibnamefont {Naletov}}, \bibinfo
		{author} {\bibfnamefont {J.}~\bibnamefont {Ben~Youssef}}, \bibinfo {author}
		{\bibfnamefont {O.}~\bibnamefont {Klein}}, \ and\ \bibinfo {author}
		{\bibfnamefont {M.}~\bibnamefont {Viret}},\ }\href {\doibase
		10.1209/0295-5075/108/57005} {\bibfield  {journal} {\bibinfo  {journal}
			{Europhys. Lett.}\ }\textbf {\bibinfo {volume} {108}},\ \bibinfo {pages}
		{57005} (\bibinfo {year} {2014})}\BibitemShut {NoStop}%
	\bibitem [{\citenamefont {Qiu}\ \emph {et~al.}(2016)\citenamefont {Qiu},
		\citenamefont {Li}, \citenamefont {Hou}, \citenamefont {Arenholz},
		\citenamefont {N\'{}Diaye}, \citenamefont {Tan}, \citenamefont {Uchida},
		\citenamefont {Sato}, \citenamefont {Okamoto}, \citenamefont {Tserkovnyak},
		\citenamefont {Qiu},\ and\ \citenamefont {Saitoh}}]{Qiu2016}%
	\BibitemOpen
	\bibfield  {author} {\bibinfo {author} {\bibfnamefont {Z.}~\bibnamefont
			{Qiu}}, \bibinfo {author} {\bibfnamefont {J.}~\bibnamefont {Li}}, \bibinfo
		{author} {\bibfnamefont {D.}~\bibnamefont {Hou}}, \bibinfo {author}
		{\bibfnamefont {E.}~\bibnamefont {Arenholz}}, \bibinfo {author}
		{\bibfnamefont {A.~T.}\ \bibnamefont {N\'{}Diaye}}, \bibinfo {author}
		{\bibfnamefont {A.}~\bibnamefont {Tan}}, \bibinfo {author} {\bibfnamefont
			{K.-i.}\ \bibnamefont {Uchida}}, \bibinfo {author} {\bibfnamefont
			{K.}~\bibnamefont {Sato}}, \bibinfo {author} {\bibfnamefont {S.}~\bibnamefont
			{Okamoto}}, \bibinfo {author} {\bibfnamefont {Y.}~\bibnamefont
			{Tserkovnyak}}, \bibinfo {author} {\bibfnamefont {Z.~Q.}\ \bibnamefont
			{Qiu}}, \ and\ \bibinfo {author} {\bibfnamefont {E.}~\bibnamefont {Saitoh}},\
	}\href {\doibase 10.1038/ncomms12670} {\bibfield  {journal} {\bibinfo
		{journal} {Nat. Commun.}\ }\textbf {\bibinfo {volume} {7}},\ \bibinfo {pages}
	{12670} (\bibinfo {year} {2016})}\BibitemShut {NoStop}%
\bibitem [{\citenamefont {Tserkovnyak}, \citenamefont {Brataas},\ and\
	\citenamefont {Bauer}(2002)}]{Tserkovnyak2002}%
\BibitemOpen
\bibfield  {author} {\bibinfo {author} {\bibfnamefont {Y.}~\bibnamefont
		{Tserkovnyak}}, \bibinfo {author} {\bibfnamefont {A.}~\bibnamefont
		{Brataas}}, \ and\ \bibinfo {author} {\bibfnamefont {G.~E.~W.}\ \bibnamefont
		{Bauer}},\ }\href {\doibase 10.1103/PhysRevB.66.224403} {\bibfield  {journal}
	{\bibinfo  {journal} {Phys. Rev. B}\ }\textbf {\bibinfo {volume} {66}},\
	\bibinfo {pages} {224403} (\bibinfo {year} {2002})}\BibitemShut {NoStop}%
\bibitem [{\citenamefont {Brataas}\ \emph {et~al.}(2012)\citenamefont
	{Brataas}, \citenamefont {Tserkovnyak}, \citenamefont {Bauer},\ and\
	\citenamefont {Kelly}}]{Brataas2012}%
\BibitemOpen
\bibfield  {author} {\bibinfo {author} {\bibfnamefont {A.}~\bibnamefont
		{Brataas}}, \bibinfo {author} {\bibfnamefont {Y.}~\bibnamefont
		{Tserkovnyak}}, \bibinfo {author} {\bibfnamefont {G.~E.~W.}\ \bibnamefont
		{Bauer}}, \ and\ \bibinfo {author} {\bibfnamefont {P.~J.}\ \bibnamefont
		{Kelly}},\ }in\ \href@noop {} {\emph {\bibinfo {booktitle} {Spin Current}}},\
\bibinfo {editor} {edited by\ \bibinfo {editor} {\bibfnamefont
		{S.}~\bibnamefont {Maekawa}}, \bibinfo {editor} {\bibfnamefont {S.~O.}\
		\bibnamefont {Valenzuela}}, \bibinfo {editor} {\bibfnamefont
		{E.}~\bibnamefont {Saitoh}}, \ and\ \bibinfo {editor} {\bibfnamefont
		{T.}~\bibnamefont {Kimura}}}\ (\bibinfo  {publisher} {Oxford University
	Pess},\ \bibinfo {address} {Oxford},\ \bibinfo {year} {2012})\ p.~\bibinfo
{pages} {87}\BibitemShut {NoStop}%
\bibitem [{\citenamefont {Rezende}\ \emph {et~al.}(2003)\citenamefont
	{Rezende}, \citenamefont {Lucena}, \citenamefont {Azevedo}, \citenamefont
	{de~Aguiar}, \citenamefont {Fermin},\ and\ \citenamefont
	{Parkin}}]{Rezende2003}%
\BibitemOpen
\bibfield  {author} {\bibinfo {author} {\bibfnamefont {S.~M.}\ \bibnamefont
		{Rezende}}, \bibinfo {author} {\bibfnamefont {M.~A.}\ \bibnamefont {Lucena}},
	\bibinfo {author} {\bibfnamefont {A.}~\bibnamefont {Azevedo}}, \bibinfo
	{author} {\bibfnamefont {F.~M.}\ \bibnamefont {de~Aguiar}}, \bibinfo {author}
	{\bibfnamefont {J.~R.}\ \bibnamefont {Fermin}}, \ and\ \bibinfo {author}
	{\bibfnamefont {S.~S.~P.}\ \bibnamefont {Parkin}},\ }\href {\doibase
	http://dx.doi.org/10.1063/1.1543126} {\bibfield  {journal} {\bibinfo
		{journal} {J. Appl. Phys.}\ }\textbf {\bibinfo {volume} {93}},\ \bibinfo
	{pages} {7717} (\bibinfo {year} {2003})}\BibitemShut {NoStop}%
\bibitem [{\citenamefont {Brataas}, \citenamefont {Bauer},\ and\ \citenamefont
	{J.}(2006)}]{Brataas2006}%
\BibitemOpen
\bibfield  {author} {\bibinfo {author} {\bibfnamefont {A.}~\bibnamefont
		{Brataas}}, \bibinfo {author} {\bibfnamefont {G.~E.~W.}\ \bibnamefont
		{Bauer}}, \ and\ \bibinfo {author} {\bibfnamefont {K.~P.}\ \bibnamefont
		{J.}},\ }\href {\doibase http://dx.doi.org/10.1016/j.physrep.2006.01.001}
{\bibfield  {journal} {\bibinfo  {journal} {Phys. Rep.}\ }\textbf {\bibinfo
		{volume} {427}},\ \bibinfo {pages} {157} (\bibinfo {year}
	{2006})}\BibitemShut {NoStop}%
\bibitem [{Tan(2014)}]{Taniguchi2014}%
\BibitemOpen
\href {\doibase 10.1103/PhysRevB.90.214407} {\bibfield  {journal} {\bibinfo
		{journal} {Phys. Rev. B}\ }\textbf {\bibinfo {volume} {90}},\ \bibinfo
	{pages} {214407} (\bibinfo {year} {2014})}\BibitemShut {NoStop}%
\bibitem [{\citenamefont {Cheng}\ \emph {et~al.}(2014)\citenamefont {Cheng},
	\citenamefont {Xiao}, \citenamefont {Niu},\ and\ \citenamefont
	{Brataas}}]{Cheng2014}%
\BibitemOpen
\bibfield  {author} {\bibinfo {author} {\bibfnamefont {R.}~\bibnamefont
		{Cheng}}, \bibinfo {author} {\bibfnamefont {J.}~\bibnamefont {Xiao}},
	\bibinfo {author} {\bibfnamefont {Q.}~\bibnamefont {Niu}}, \ and\ \bibinfo
	{author} {\bibfnamefont {A.}~\bibnamefont {Brataas}},\ }\href {\doibase
	10.1103/PhysRevLett.113.057601} {\bibfield  {journal} {\bibinfo  {journal}
		{Phys. Rev. Lett.}\ }\textbf {\bibinfo {volume} {113}},\ \bibinfo {pages}
	{057601} (\bibinfo {year} {2014})}\BibitemShut {NoStop}%
\bibitem [{\citenamefont {Zhang}\ \emph {et~al.}(2014)\citenamefont {Zhang},
	\citenamefont {Jungfleisch}, \citenamefont {Jiang}, \citenamefont {Pearson},
	\citenamefont {Hoffmann}, \citenamefont {Freimuth},\ and\ \citenamefont
	{Mokrousov}}]{Zhang2014}%
\BibitemOpen
\bibfield  {author} {\bibinfo {author} {\bibfnamefont {W.}~\bibnamefont
		{Zhang}}, \bibinfo {author} {\bibfnamefont {M.~B.}\ \bibnamefont
		{Jungfleisch}}, \bibinfo {author} {\bibfnamefont {W.}~\bibnamefont {Jiang}},
	\bibinfo {author} {\bibfnamefont {J.~E.}\ \bibnamefont {Pearson}}, \bibinfo
	{author} {\bibfnamefont {A.}~\bibnamefont {Hoffmann}}, \bibinfo {author}
	{\bibfnamefont {F.}~\bibnamefont {Freimuth}}, \ and\ \bibinfo {author}
	{\bibfnamefont {Y.}~\bibnamefont {Mokrousov}},\ }\href {\doibase
	10.1103/PhysRevLett.113.196602} {\bibfield  {journal} {\bibinfo  {journal}
		{Phys. Rev. Lett.}\ }\textbf {\bibinfo {volume} {113}},\ \bibinfo {pages}
	{196602} (\bibinfo {year} {2014})}\BibitemShut {NoStop}%
\bibitem [{\citenamefont {Kravets}\ \emph {et~al.}(2012)\citenamefont
	{Kravets}, \citenamefont {Timoshevskii}, \citenamefont {Yanchitsky},
	\citenamefont {Bergmann}, \citenamefont {Buhler}, \citenamefont {Andersson},\
	and\ \citenamefont {Korenivski}}]{Kravets2012}%
\BibitemOpen
\bibfield  {author} {\bibinfo {author} {\bibfnamefont {A.~F.}\ \bibnamefont
		{Kravets}}, \bibinfo {author} {\bibfnamefont {A.~N.}\ \bibnamefont
		{Timoshevskii}}, \bibinfo {author} {\bibfnamefont {B.~Z.}\ \bibnamefont
		{Yanchitsky}}, \bibinfo {author} {\bibfnamefont {M.~A.}\ \bibnamefont
		{Bergmann}}, \bibinfo {author} {\bibfnamefont {J.}~\bibnamefont {Buhler}},
	\bibinfo {author} {\bibfnamefont {S.}~\bibnamefont {Andersson}}, \ and\
	\bibinfo {author} {\bibfnamefont {V.}~\bibnamefont {Korenivski}},\ }\href
{\doibase 10.1103/PhysRevB.86.214413} {\bibfield  {journal} {\bibinfo
		{journal} {Phys. Rev. B}\ }\textbf {\bibinfo {volume} {86}},\ \bibinfo
	{pages} {214413} (\bibinfo {year} {2012})}\BibitemShut {NoStop}%
\bibitem [{\citenamefont {Tserkovnyak}\ \emph {et~al.}(2005)\citenamefont
	{Tserkovnyak}, \citenamefont {Brataas}, \citenamefont {Bauer},\ and\
	\citenamefont {Halperin}}]{Tserkovnyak2005}%
\BibitemOpen
\bibfield  {author} {\bibinfo {author} {\bibfnamefont {Y.}~\bibnamefont
		{Tserkovnyak}}, \bibinfo {author} {\bibfnamefont {A.}~\bibnamefont
		{Brataas}}, \bibinfo {author} {\bibfnamefont {G.~E.~W.}\ \bibnamefont
		{Bauer}}, \ and\ \bibinfo {author} {\bibfnamefont {B.~I.}\ \bibnamefont
		{Halperin}},\ }\href {\doibase 10.1103/RevModPhys.77.1375} {\bibfield
	{journal} {\bibinfo  {journal} {Rev. Mod. Phys.}\ }\textbf {\bibinfo {volume}
		{77}},\ \bibinfo {pages} {1375} (\bibinfo {year} {2005})}\BibitemShut
{NoStop}%
\bibitem [{\citenamefont {Baryakhtar}\ and\ \citenamefont
	{Ivanov}(1979)}]{Baryakhtar1979}%
\BibitemOpen
\bibfield  {author} {\bibinfo {author} {\bibfnamefont {I.~V.}\ \bibnamefont
		{Baryakhtar}}\ and\ \bibinfo {author} {\bibfnamefont {B.~A.}\ \bibnamefont
		{Ivanov}},\ }\href@noop {} {\bibfield  {journal} {\bibinfo  {journal} {Sov.
			J. Low Temp. Phys.}\ }\textbf {\bibinfo {volume} {5}},\ \bibinfo {pages}
	{361} (\bibinfo {year} {1979})}\BibitemShut {NoStop}%
\bibitem [{\citenamefont {Gomonay}\ and\ \citenamefont
	{Loktev}(2008)}]{Gomonay2008}%
\BibitemOpen
\bibfield  {author} {\bibinfo {author} {\bibfnamefont {H.}~\bibnamefont
		{Gomonay}}\ and\ \bibinfo {author} {\bibfnamefont {V.}~\bibnamefont
		{Loktev}},\ }\href {\doibase 10.3379/msjmag.32.535} {\bibfield  {journal}
	{\bibinfo  {journal} {J. Magn. Soc. Jpn.}\ }\textbf {\bibinfo {volume}
		{32}},\ \bibinfo {pages} {535} (\bibinfo {year} {2008})}\BibitemShut
{NoStop}%
\bibitem [{\citenamefont {Callen}(1948)}]{Callen1948}%
\BibitemOpen
\bibfield  {author} {\bibinfo {author} {\bibfnamefont {H.~B.}\ \bibnamefont
		{Callen}},\ }\href {\doibase 10.1103/PhysRev.73.1349} {\bibfield  {journal}
	{\bibinfo  {journal} {Phys. Rev.}\ }\textbf {\bibinfo {volume} {73}},\
	\bibinfo {pages} {1349} (\bibinfo {year} {1948})}\BibitemShut {NoStop}%
\bibitem [{\citenamefont {Cheng}\ and\ \citenamefont {Niu}(2014)}]{Cheng2014a}%
\BibitemOpen
\bibfield  {author} {\bibinfo {author} {\bibfnamefont {R.}~\bibnamefont
		{Cheng}}\ and\ \bibinfo {author} {\bibfnamefont {Q.}~\bibnamefont {Niu}},\
}\href {\doibase 10.1103/PhysRevB.89.081105} {\bibfield  {journal} {\bibinfo
	{journal} {Phys. Rev. B}\ }\textbf {\bibinfo {volume} {89}},\ \bibinfo
{pages} {081105} (\bibinfo {year} {2014})}\BibitemShut {NoStop}%
\bibitem [{\citenamefont {Brataas}, \citenamefont {Tserkovnyak},\ and\
	\citenamefont {Bauer}(2011)}]{Brataas2011}%
\BibitemOpen
\bibfield  {author} {\bibinfo {author} {\bibfnamefont {A.}~\bibnamefont
		{Brataas}}, \bibinfo {author} {\bibfnamefont {Y.}~\bibnamefont
		{Tserkovnyak}}, \ and\ \bibinfo {author} {\bibfnamefont {G.~E.~W.}\
		\bibnamefont {Bauer}},\ }\href {\doibase 10.1103/PhysRevB.84.054416}
{\bibfield  {journal} {\bibinfo  {journal} {Phys. Rev. B}\ }\textbf {\bibinfo
		{volume} {84}},\ \bibinfo {pages} {054416} (\bibinfo {year}
	{2011})}\BibitemShut {NoStop}%
\bibitem [{\citenamefont {Ohnuma}\ \emph {et~al.}(2014)\citenamefont {Ohnuma},
	\citenamefont {Adachi}, \citenamefont {Saitoh},\ and\ \citenamefont
	{Maekawa}}]{Ohnuma2014}%
\BibitemOpen
\bibfield  {author} {\bibinfo {author} {\bibfnamefont {Y.}~\bibnamefont
		{Ohnuma}}, \bibinfo {author} {\bibfnamefont {H.}~\bibnamefont {Adachi}},
	\bibinfo {author} {\bibfnamefont {E.}~\bibnamefont {Saitoh}}, \ and\ \bibinfo
	{author} {\bibfnamefont {S.}~\bibnamefont {Maekawa}},\ }\href {\doibase
	10.1103/PhysRevB.89.174417} {\bibfield  {journal} {\bibinfo  {journal} {Phys.
			Rev. B}\ }\textbf {\bibinfo {volume} {89}},\ \bibinfo {pages} {174417}
	(\bibinfo {year} {2014})}\BibitemShut {NoStop}%
\bibitem [{\citenamefont {Acharyya}\ \emph {et~al.}(2011)\citenamefont
	{Acharyya}, \citenamefont {Nguyen}, \citenamefont {Pratt},\ and\
	\citenamefont {Bass}}]{Acharyya2011}%
\BibitemOpen
\bibfield  {author} {\bibinfo {author} {\bibfnamefont {R.}~\bibnamefont
		{Acharyya}}, \bibinfo {author} {\bibfnamefont {H.~Y.~T.}\ \bibnamefont
		{Nguyen}}, \bibinfo {author} {\bibfnamefont {W.~P.}\ \bibnamefont {Pratt}}, \
	and\ \bibinfo {author} {\bibfnamefont {J.}~\bibnamefont {Bass}},\ }\href
{\doibase 10.1063/1.3535340} {\bibfield  {journal} {\bibinfo  {journal} {J.
			Appl. Phys.}\ }\textbf {\bibinfo {volume} {109}},\ \bibinfo {pages} {07C503}
	(\bibinfo {year} {2011})}\BibitemShut {NoStop}%
\bibitem [{\citenamefont {Mizukami}, \citenamefont {Ando},\ and\ \citenamefont
	{Miyazaki}()}]{Mizukami2001a}%
\BibitemOpen
\bibfield  {author} {\bibinfo {author} {\bibfnamefont {S.}~\bibnamefont
		{Mizukami}}, \bibinfo {author} {\bibfnamefont {Y.}~\bibnamefont {Ando}}, \
	and\ \bibinfo {author} {\bibfnamefont {T.}~\bibnamefont {Miyazaki}},\ }\href
{http://stacks.iop.org/1347-4065/40/i=2R/a=580} {\bibfield  {journal}
	{\bibinfo  {journal} {Jpn. J. Appl. Phys.}\ }\textbf {\bibinfo {volume}
		{40}},\ \bibinfo {pages} {580}}\BibitemShut {NoStop}%
\bibitem [{\citenamefont {Kravets}\ \emph {et~al.}(2016)\citenamefont
	{Kravets}, \citenamefont {Polishchuk}, \citenamefont {Dzhezherya},
	\citenamefont {Tovstolytkin}, \citenamefont {Golub},\ and\ \citenamefont
	{Korenivski}}]{Kravets2016}%
\BibitemOpen
\bibfield  {author} {\bibinfo {author} {\bibfnamefont {A.~F.}\ \bibnamefont
		{Kravets}}, \bibinfo {author} {\bibfnamefont {D.~M.}\ \bibnamefont
		{Polishchuk}}, \bibinfo {author} {\bibfnamefont {Y.~I.}\ \bibnamefont
		{Dzhezherya}}, \bibinfo {author} {\bibfnamefont {A.~I.}\ \bibnamefont
		{Tovstolytkin}}, \bibinfo {author} {\bibfnamefont {V.~O.}\ \bibnamefont
		{Golub}}, \ and\ \bibinfo {author} {\bibfnamefont {V.}~\bibnamefont
		{Korenivski}},\ }\href {\doibase 10.1103/PhysRevB.94.064429} {\bibfield
	{journal} {\bibinfo  {journal} {Phys. Rev. B}\ }\textbf {\bibinfo {volume}
		{94}},\ \bibinfo {pages} {064429} (\bibinfo {year} {2016})}\BibitemShut
{NoStop}%
\bibitem [{\citenamefont {Montoya}\ \emph {et~al.}(2016)\citenamefont
	{Montoya}, \citenamefont {Omelchenko}, \citenamefont {Coutts}, \citenamefont
	{Lee-Hone}, \citenamefont {H\"ubner}, \citenamefont {Broun}, \citenamefont
	{Heinrich},\ and\ \citenamefont {Girt}}]{Montoya2016}%
\BibitemOpen
\bibfield  {author} {\bibinfo {author} {\bibfnamefont {E.}~\bibnamefont
		{Montoya}}, \bibinfo {author} {\bibfnamefont {P.}~\bibnamefont {Omelchenko}},
	\bibinfo {author} {\bibfnamefont {C.}~\bibnamefont {Coutts}}, \bibinfo
	{author} {\bibfnamefont {N.~R.}\ \bibnamefont {Lee-Hone}}, \bibinfo {author}
	{\bibfnamefont {R.}~\bibnamefont {H\"ubner}}, \bibinfo {author}
	{\bibfnamefont {D.}~\bibnamefont {Broun}}, \bibinfo {author} {\bibfnamefont
		{B.}~\bibnamefont {Heinrich}}, \ and\ \bibinfo {author} {\bibfnamefont
		{E.}~\bibnamefont {Girt}},\ }\href {\doibase 10.1103/PhysRevB.94.054416}
{\bibfield  {journal} {\bibinfo  {journal} {Phys. Rev. B}\ }\textbf {\bibinfo
		{volume} {94}},\ \bibinfo {pages} {054416} (\bibinfo {year}
	{2016})}\BibitemShut {NoStop}%
\end{thebibliography}
%

\end{document}